\documentstyle[prb,twocolumn,aps,rotate,epsf,epsfig,floats]{revtex}

\def\rbx#1{\raisebox{-0.3ex}{$\scriptstyle #1$}}

\begin{document}
\setlength{\unitlength}{1cm}
\renewcommand{\arraystretch}{1.4}

\title{Magneto-elastic excitations in spin-Peierls systems}

\author{Michael Holicki\thanks{Present address: Mathematisches Institut, Universit\"at Leipzig, D-04109 Leipzig, Germany} and Holger Fehske}
\address{Physics Department, University of Bayreuth, D-95440
Bayreuth, Germany} 

\author{Ralph Werner}
\address{Physics Department, University of Wuppertal, D-42097
Wuppertal, Germany, and \\ Physics Department, Brookhaven National
Laboratory, Upton, NY 11973-5000} 

\date{\today}

\maketitle

\centerline{Preprint. Typeset using REV\TeX}

\begin{abstract}
Within the random phase approximation to the spin-Peierls transition
two parameter regimes of phonon softening and hardening are
present. Magneto-elastic excitations are discussed in detail for
phonons coupled to the exactly solvable model of XY spin chains for
both regimes, leading to a modified interpretation of the 30
cm$^{-1}$ mode in CuGeO$_3$. Frustrated Heisenberg chains
coupled to phonons satisfactory describe the pre-transitional
quasi-elastic scattering in CuGeO$_3$. A real space interpretation of
the quasi-elastic scattering is given justifying effective Ising model
approaches.
\end{abstract}
\pacs{PACS numbers: 64.70.-p, 63.20.Kr, 75.10 Jm}


\section{Introduction}

The combined approach of the random-phase-approximation (RPA) for the
spin-phonon coupling and bosonization for the spin dynamics applied 
by Cross and Fisher\cite{CF79} to describe the spin Peierls-transition is
consistent not only with the phonon softening in materials as TTFCuBDT
as initially believed but also with the hardening of the Peierls-active
phonon modes in CuGeO$_3$.\cite{Gros98CGO,Braden98CGO} The
applicability of RPA is supported by the good agreement of mean-field
results with experiments\cite{Kluemper99CGO,Werner99CGO,BKW99} and the
Ginzburg criterion.\cite{Werner99CGO,BKW99,BNR98}  

In the parameter regime where phonon hardening occurs the
RPA calculations predict the appearance of spectral weight in the
center of the phonon spectrum ($\omega\sim 0$) as a precursor of the
phase transition. The precursor has been observed experimentally in
CuGeO$_3$ and its temperature dependent intensity in
neutron\cite{BradenHabil,Hirota95CGO} and
X-ray\cite{Pouget96CGO,WKC+00} scattering experiments has been shown
to be satisfactorily reproduced within RPA.\cite{Werner00ED} We
discuss in this paper the details of the precursor such as its
momentum space dependence,\cite{Hirota95CGO,BradenHabil,WKC+00}
the extracted correlation lengths,\cite{Hirota95CGO,Pouget96CGO,WKC+00}
its frequency dependence, and its real-space
interpretation.

The relevant magnetic correlation function in the RPA approach is the
dynamic dimer-dimer correlation
function.\cite{CF79,Gros98CGO,WernerThesis} The determination of  
dynamic correlation functions is not evident even for exactly
solvable one-dimensional models.\cite{Betheansatz,FLS97} The XY
model is an exception\cite{Fradkin} and since there are similarities
to the Heisenberg model\cite{YMV96,Werner00ED} it is an appropriate
model to derive qualitative results exactly. 
Our studies of the coupling of phonons to XY spin chains show that the
quasi-elastic scattering is the precursor of a new magneto-elastic
excitation appearing at the phase transition. In the regime of phonon
softening the mixed magneto-elastic nature of the ``soft phonon'' also
becomes apparent.   

The spin-phonon coupled Hamiltonian $H = H_s + H_{p} + H_{sp}$
relevant for spin-Peierls systems has been derived explicitly for
CuGeO$_3$.\cite{Werner99CGO} It consists of three parts. One is the
Heisenberg spin-chain Hamiltonian 
\begin{equation}\label{Hs}
H_s=J\ \sum_{\bf l}\ 
      {\bf S}_{\bf l}\cdot{\bf S}_{{\bf l}+\hat{z}}
    + J_2\ \sum_{\bf l}\ {\bf S}_{\bf l}\cdot{\bf S}_{{\bf l}+2\hat{z}}
\end{equation}
with the superexchange integrals $J$ and $J_2$ between nearest-neighbor
(NN) and next-nearest-neighbor (NNN) Cu $d$ orbitals, respectively,
and spin 1/2 operators ${\bf S}_{\bf l}$ at Cu site ${\bf l}$ in the
three-dimensional lattice. $\hat{z}$ is a unit vector along the
spin-chain direction. 
The harmonic phonon part
\begin{equation}\label{Hpbose}
H_p=\sum_{{\bf q},\nu}\hbar\Omega_{\nu,{\bf q}}
\left(b^\dagger_{\nu,{\bf q}}b^{}_{\nu,{\bf
q}}+\frac{1}{2}\right)
\end{equation}
contains the dispersions $\Omega_{\nu,{\bf q}}$ for the relevant
phonon modes\cite{Braden98CGO} labeled $\nu\in\{1,2,3,4\}$ and Bose
operators $b^\dagger_{\nu,{\bf q}}$ and $b^{}_{\nu,{\bf q}}$. 
The spin-phonon coupling term is given by 
\begin{equation}\label{Hspbose}
H_{sp}=\frac{1}{\sqrt{N}}\sum_{\bf q}\ Y_{-{\bf q}}\ \sum_\nu\
                g_{\nu,{\bf q}}\
\left(b^\dagger_{\nu,-{\bf q}}+b^{}_{\nu,{\bf q}}\right).
\end{equation}
$N$ is the number of unit cells in the lattice.
The coupling constants $g_{\nu,{\bf q}}$ depend on the polarization 
vectors of phonon mode $\nu$ and the Fourier transformed dimer
operator is defined as 
\begin{equation}\label{defYq}
Y_{-{\bf q}} := \sum_{\bf l}\ e^{i{\bf q}{\bf R}_{\bf l}}\ 
        {\bf S}_{\bf l}\cdot{\bf S}_{{\bf l}+\hat{z}}.
\end{equation}

To compare with neutron or X-ray scattering data the phonon dynamic
structure factor $S({\bf q},\omega)$ has to be determined. It is given
via the imaginary part of the retarded normal coordinate propagator 
$D^{\rm ret}_\nu({\bf q},\omega)$
\begin{equation} 
S({\bf q},\omega) = -\frac{1}{\pi} \frac{
    \sum_\nu {\rm Im} D^{\rm ret}_\nu({\bf q},\omega)} 
{1-\exp(-\beta\hbar\omega)}\,.
\label{dynamical}
\end{equation}
We introduced the inverse temperature $\beta=1/(k_{\rm B}T)$. The
retarded normal coordinate propagator is obtained through analytical
continuation of the Matsubara propagator onto the real frequency axis 
\begin{equation}\label{cont}
D^{\rm ret}_\nu({\bf q},\omega)=\lim_{\epsilon\to 0}
       D_\nu({\bf q},i\omega_n \to \hbar\omega + i\epsilon)\,, 
\end{equation}
where the latter is given in RPA \cite{WernerThesis} by
\begin{eqnarray}\label{DMatsubara}
D_{\nu}({\bf q},i\omega_n) &=& D^{(0)}_{\nu}({\bf q},i\omega_n)
\nonumber\\[1ex]&&\hspace{-8ex}\times\
\frac{1- \chi({\bf q},i\omega_n) 
         \sum\limits_{\nu'\neq\nu}
             g_{\nu',{\bf q}}g_{\nu',-{\bf q}}\
                    D^{(0)}_{\nu'}({\bf q},i\omega_n)} 
	{1- \chi({\bf q},i\omega_n)\ 
         \sum\limits_{\nu'}\ g_{\nu',{\bf q}}g_{\nu',-{\bf q}}\
                    D^{(0)}_{\nu'}({\bf q},i\omega_n)} 
\end{eqnarray}
with bosonic Matsubara frequencies $\omega_n=2\pi n/\beta$.
The unperturbed propagator is
\begin{equation}\label{Dnullretarded}
D^{(0)}_{\nu}({\bf q},i\omega_n) = 
         - \frac{2\ \hbar\Omega_{\nu,{\bf q}}}
         {\omega_n^2 + \hbar^2\Omega^2_{\nu,{\bf q}}}\,.
\end{equation}

The dimer-dimer correlation function 
\begin{equation}\label{chi0}
\chi(q_z,i\omega_n) = 
- \frac{1}{N}\int_0^\beta d\tau
     \ {\rm e}^{i \omega_n \tau}\
     \left\langle Y_{\bf q}(\tau) Y_{-\bf q}(0) \right\rangle\
\end{equation}
depends only on momenta $q_z$ along the spin chains. Since the exact
determination of $\chi(q_z,i\omega_n)$ in the case of the Heisenberg
model is impossible we first study the case of the XY model. In Sec.\
\ref{Quasielastic} we then turn to the application of the RPA results
on CuGeO$_3$, where a combination of analytical and numerical results
is used to determine the dimer-dimer correlation function as accurate
as possible.


 \section{Magnetostrictive XY model}\label{minimal}

It is commonly accepted that the basic features of 
the spin-Peierls transition are well described by 
a one-dimensional spin model coupled 
magneto-elastically to the three-dimensional phonon system.
The neglect of magnetic inter-chain coupling and frustration
effects is certainly justified if the spin-phonon interaction  
dominates these spin interactions and causes the dimerization.
Concerning the spin system, Caron and Moukouri\cite{CaronMoukouri96} 
showed that the simple XY spin chain model with $J_2=0$, 
\begin{equation}\label{HsXY}
H_s=J\ \sum_{l\equiv l_z=1}^N
             \left( S^x_{l} S^x_{l+1}
                  +S^y_{l} S^y_{l+1} \right),
\end{equation}
contains the relevant physics of a spin-Peierls system,
mainly because its excitation spectrum exhibits the requisite 
degeneracy with the ground state.\cite{Brayetal} 
In general, the mixed dimensionality of magnetic and spin-phonon
interactions makes a theoretical treatment difficult.
However, within the RPA (mean-field) approach in the spin-phonon 
coupling results obtained for a purely 1D phonon system will be 
the same as those for a 3D phonon system,\cite{techcomment}
whose polarization vectors satisfy 
${\bf e}_{\nu,q_z\cdot\hat{z}}\hat{z}=\delta_{\nu,1}$  
and whose dispersion along the chain is
\begin{equation}\label{1Ddisp} 
\Omega^2_{q}\equiv\Omega^2_{1,q\cdot\hat{z}}=\frac{1}{2}\Omega_\pi^2
\left[1-\cos(q)\right],
\end{equation}
i.e., only one acoustic phonon branch couples to the magnetic system.
In Eq.~(\ref{1Ddisp}), the longitudinal wave number $q=q_z$ 
is given in units of the reciprocal lattice spacing $1/c$.  
\subsection{Uniform Phase}
In solving the magnetostrictive XY model it is convenient to 
transform the spin-operators ($S^x, S^y$) to operators of spinless 
fermions ($d^{(\dagger)}_l $),   
via the Jordan-Wigner transformation.\cite{Mahan}
Then, in the uniform phase above $T_{\rm SP}$, we start from the following
(Fourier transformed) Hamiltonian 
\begin{eqnarray}\label{Hminimal}
H&=&H_p+H_{s}+H_{sp}\nonumber\\
&=&\sum_q\Omega_qb^\dagger_qb_q+\sum_kE_kd^\dagger_kd_k\nonumber\\
& &+\frac{1}{\sqrt{N}}\sum_qg_qY_{-q}(b_q+b^\dagger_{-q})
\end{eqnarray}
with 
\begin{eqnarray}
E_k&=&\cos(k),\\
g_q&=&\left(\frac{\lambda\Omega_\pi^2\pi}{\Omega_q}\right)^{\frac{1}{2}}
\left(1-e^{iq}\right),\\
\lambda&=&\frac{g^2}{2\pi m\Omega_\pi^2},\ \ \ \ g=\left.\frac{dJ}{dr}
\right|_{r=c},
\end{eqnarray}
and $q, k \in ]-\pi,\pi]$ (in this section, we drop the factors 
$\hbar$, $k_B$ and define all energies 
in units of $J$).  
The dimer operator (\ref{defYq}) is now
\begin{eqnarray}\label{defYqXY}
Y_{-q} &=& \sum_{l}\ e^{iql}\ \left( S^x_{l} S^x_{l+1}+
S^y_{l} S^y_{l+1} \right)\nonumber\\
&=&\frac{1}{2}\sum_k\left(e^{i(k-q)}+e^{-ik}\right)d^\dagger_kd_{k-q}.
\end{eqnarray}

Inserting Eq.~(\ref{1Ddisp}) we see that the ground-state 
and thermodynamic properties of the model~(\ref{Hminimal}) 
are governed by two independent control parameters: 
(i) the dimensionless coupling constant $\lambda$ and 
(ii) the ratio of phononic to magnetic energy scale $\Omega_\pi$. 
$\lambda$ is independent of the ion mass  
$m$ because $\Omega_\pi^2\sim 1/m$. It will turn out that the 
transition temperature and the static dimerization is a 
function of $\lambda$ alone. For a constant $\lambda$, 
$\Omega_\pi$ is a measure for the mass of the ions; 
small values of $\Omega_\pi$ describe the adiabatic, 
large values the anti-adiabatic regime.

For the  model~(\ref{Hminimal}), the RPA Matsubara 
Green's function~(\ref{DMatsubara}) becomes 
\begin{equation}
D(q,i\omega_n)=D^{(0)}(q,i\omega_n)\frac{1}{1-D^{(0)}(q,i\omega_n) 
P(q,i\omega_n)},
\end{equation}
where the self energy is defined as
\begin{equation}
P(q,i\omega_n)=g_qg_{-q}\chi(q,i\omega_n).
\end{equation}
The dimer-dimer correlation function (\ref{chi0}) is
\begin{equation}
\label{dimdimcor}
\chi(q,i\omega_n)=\frac{1}{4\pi}\int_{-\pi}^{\pi}dk\left(1+\cos(2k+q)\right)
K_{k,q}(i\omega_n),
\end{equation}
with the Lindhard kernel 
\begin{eqnarray}\label{Lindhard1}
K_{k,q}(i\omega_n)&=&\frac{f_{k+q}-f_{k}}{i\omega_n+ E_{k+q}- E_k},
\end{eqnarray}
where $f_k=1/(e^{\beta E_k}+1)$ is the Fermi distribution function.

\begin{figure}[bt]
\begin{center}
\epsfxsize=0.4\textwidth
\epsffile{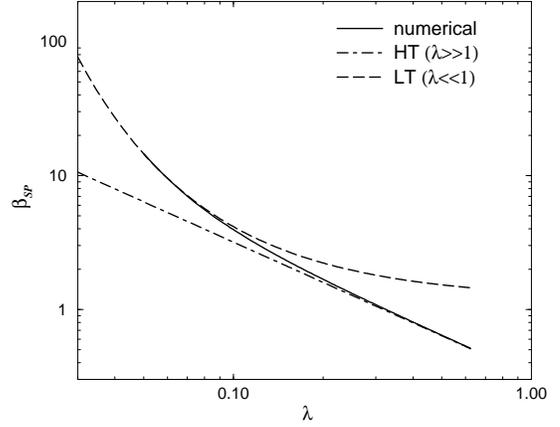}
\caption{\sl Inverse spin-Peierls transition temperature $\beta_{\rm SP}$ vs. 
coupling constant $\lambda$ (solid line). The dot-dashed (dashed) curve
denotes the high (low) temperature approximation.}
\label{TSPplot}
\end{center}
\end{figure}

Having calculated the Matsubara propagator we can easily obtain the retarded 
Green's function on the real frequency axis according to Eq.~(\ref{cont}). 
A structural instability is always 
connected to a pole of $D^{\rm ret}(q,\omega)$ at $\omega=0$, leading 
to a spontaneous transition to a broken-symmetry ground state at $T=0$. 
At finite temperature, the instability condition for a lattice 
distortion with wave number $q$ is
\begin{eqnarray}\label{softcond}
\frac{1}{2\lambda}&=&\int_{-\pi}^\pi dk(1+\cos(2k+q))K_{k,q}(0),
\end{eqnarray}
which is the same result as derived by Lima and Tsallis\cite{LimaTsallis81} 
in the adiabatic limit ($m\to\infty$). It turns out that the $\pi$ mode is the 
first (and only) one that gets unstable. Therefore the lattice will dimerize 
below a transition temperature $T_{\rm SP}$, which can be obtained 
from the numerical solution of Eq.~(\ref{softcond}) at $q=\pi$. 
The inverse transition temperature is shown in Fig.~\ref{TSPplot} 
together with the results obtained from  high and low temperature
expansions,
\begin{eqnarray}
\beta_{\rm SP}&=&\frac{1}{\lambda\pi}\hspace*{2.7cm}\lambda\gg 1,\\
\beta_{\rm SP}&=& 1.19\cdot\exp\left(\frac{1}{8\lambda}\right)\qquad\lambda\ll 1,
\end{eqnarray}
respectively.
\subsection{Dimerized Phase}
\begin{figure}[bt]
\begin{center}
\epsfxsize=0.48\textwidth
\epsffile{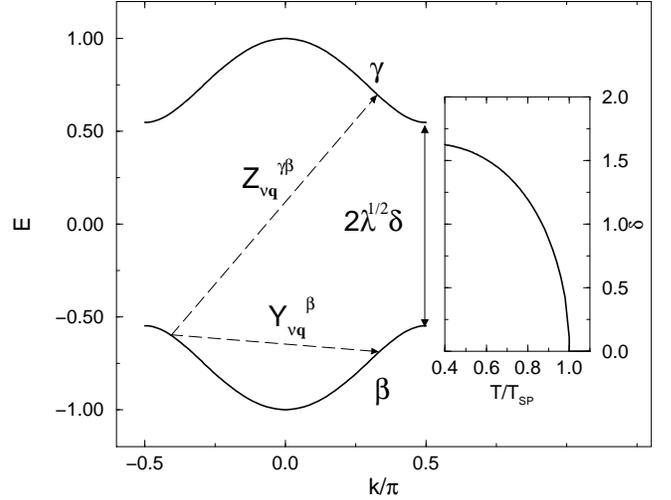}
\caption{\sl Fermion band structure in the dimerized phase (left panel). 
The dashed arrows indicate the processes described by the $Y$ and $Z$ 
operators. The right panel shows the temperature dependence 
of the dimerization $\delta$ 
for $\lambda=0.1$.}\label{ordparplot}
\end{center}
\end{figure}

Below the transition temperature $T_{\rm SP}$ the systems is in a
less-symmetric but lower energy configuration. The lattice is dimerized 
which causes the unit cell to double in size, which in reciprocal 
space means that $q, k \in ]-\pi/2,\pi/2]$. 
To include a static dimerization $\delta$ in our Hamiltonian explicitely, 
we perform a unitary transformation 
\begin{equation}
\tilde{H}=e^SHe^{-S},
\end{equation}
with 
\begin{equation}
S=\frac{1}{4}\sqrt{\frac{N}{\pi\Omega_\pi}}\delta(b_\pi-b^\dagger_{\pi}).
\end{equation}
In a next step, we redefine the Fourier transformed fermion
operators\cite{BP72}
\begin{eqnarray}
c_k&=&\sqrt{\frac{2}{N}}\sum_{l=1}^{\frac{N}{2}}e^{-i(2l-1)k}d_{2l-1},\\
\bar{c}_k&=&\sqrt{\frac{2}{N}}\sum_{l=1}^{\frac{N}{2}}e^{-i(2l)k}d_{2l},
\end{eqnarray}
and remove the non-diagonal $c$-$\bar{c}$ cross-terms 
by a canonical Bogoliubov  transformation
\begin{equation}
c_k=\frac{1}{\sqrt{2}}(\gamma_k+\beta_k),\hspace{1cm}\bar{c}_k
=\frac{1}{\sqrt{2}}(\gamma_k-\beta_k)e^{i\Theta_k},
\end{equation}
where
\begin{equation}
\Theta_k=\arctan\left(\sqrt{\lambda}\delta\right).
\end{equation}
For the phonon part, we now have two modes denoted by $\nu$, an acoustical 
($\nu=0$) and an optical ($\nu=1$),
where in the reduced Brillouin zone the phonon operators
are 
\begin{equation}
b_{\nu,q}=\left\{ \begin{array}{r@{\qquad\qquad}l}
b_{q+\nu\pi}&q\leq 0\\
b_{q-\nu\pi}&q> 0\;,
\end{array}\right.
\end{equation}
and  the dispersion is 
\begin{equation}
\Omega_{\nu,q}=\Omega_{q+\nu\pi}.
\end{equation}
Finally the Hamiltonian describing the dimerized phase takes the form
\begin{eqnarray}
\tilde{H}&=& {H}_{p}+ {H}_{s}+ {H}_{dp}+ {H}_{elast}+ {H}_{sp}\nonumber\\
&=&\sum_{q,\nu}{\Omega}_{\nu,q}b^\dagger_{\nu,q}b_{\nu,q}+
\sum_k E_k(\gamma^\dagger_k\gamma_k-\beta^\dagger_k\beta_k)\nonumber\\
& &-\delta\sqrt{\frac{N\Omega_\pi}{16\pi}}(b_{1,0}
+b^\dagger_{1,0})+
\frac{N\delta^2}{16\pi}\nonumber\\
& &+\frac{1}{\sqrt{N}}\sum_{\nu,q}g_{\nu,q}(Y^\beta_{\nu,-q}-Y^\gamma_{\nu,-q}-
Z^{\gamma\beta}_{\nu,-q}+Z^{\beta\gamma}_{\nu,-q})\nonumber\\
& &\hspace{3cm}\times(b_{\nu,q}+b^\dagger_{\nu,-q}),
\end{eqnarray}
with the operators
\begin{eqnarray}
Y^\gamma_{\nu,-q}&=&\sum_k\left((-1)^\nu e^{i(k-q)}+e^{-ik}\right)
T_{k,\nu,q}\gamma^\dagger_k\gamma_{k-q},\\
Z^{\gamma\beta}_{\nu,-q}&=&\sum_k\left((-1)^\nu e^{i(k-q)}+e^{-ik}\right)
T_{k,\nu+1,q}\gamma^\dagger_k\beta_{k-q},
\end{eqnarray}
and
\begin{eqnarray}
g_{\nu,q}&=&\left(\frac{\lambda\Omega_\pi^2\pi}{\Omega_{\nu,q}}\right)
^{\frac{1}{2}}\left(1-(-1)^\nu e^{iq}\right),\\
E_k&=&\sqrt{\cos^2(k)+\lambda\delta^2\sin^2(k)}= E^\gamma_k=- E^\beta_k,\\
T_{k,\nu,q}&=&\frac{1}{4}\left(e^{i\Theta_{k-q}}+\alpha_{k-q}(-1)^\nu 
e^{-i\Theta_k}\right).
\end{eqnarray}
The phase factor  $\alpha_{k-q}$ is $1$ for normal  
($|k-q|<\frac{\pi}{2}$) and $-1$ for Umklapp 
($|k-q|>\frac{\pi}{2}$)
processes.
Accordingly, instead of one fermion band as in the uniform case, 
we now have two bands, separated by a gap proportional to $\delta$. 
The operators $Y$ and $Z$ describe intra- and inter-band transitions, 
respectively (see Fig. \ref{ordparplot}).

The parameter $\delta$ is not yet defined. 
Due to the invariance of the trace 
under canonical transformations the following 
relation holds for all values of $\delta$
\begin{eqnarray}
\label{equcon}
0&=&-\frac{\partial}{\partial\delta}\frac{1}{\beta}\ln\left(\textrm{Tr}
e^{-\beta \tilde{H}}\right)\nonumber\\
&=&\sum_k\frac{\lambda\delta}{E_k}\langle\gamma_k^\dagger\gamma_k-
\beta^\dagger_k\beta_k\rangle_{\tilde{H}}-\sqrt{\frac{N\Omega_\pi}{16\pi}}
\langle b_{1,0}^{}+b^\dagger_{1,0}\rangle_{\tilde{H}}\nonumber\\
&&+\frac{N\delta}{8\pi}+\left\langle \frac{\partial}{\partial\delta}
H_{sp}\right\rangle_{\tilde{H}}.
\end{eqnarray}
To determine $\delta$, we demand that the phonon 
coordinates have expectation value zero with respect to $\tilde{H}$:
\begin{equation}
\langle b_{1,0}+b^\dagger_{1,0}
\rangle_{\tilde{H}}\stackrel{!}{=}0.
\end{equation} 
This means that we generated a Hamiltonian $\tilde{H}$ with
$q=0,\ \nu=1$ modes shifted by $\delta$ in a way that the expectation
value of the $b$-operators under the new Hamiltonian is zero. Therefore
$\delta$ is the equilibrium position of the $\pi$ mode under the original
Hamiltonian $H$ (the static dimerization): 
\begin{equation}
\delta\propto\langle b_\pi+b^\dagger_\pi\rangle_H.
\end{equation} 

If the calculations are not done in Fourier but in real space, 
it can easily be shown\cite{HolickiDiploma} that the $\delta$ determined by 
this prescription is directly proportional to the magnetic order parameter 
used by other authors\cite{Wellein98}: 
\begin{equation}
\delta\propto \left\langle \frac{1}{N}
\sum_l(-1)^l(S_l^xS_{l+1}^x+S_l^yS_{l+1}^y)\right\rangle_{\tilde{H}}.
\end{equation}
From Eq.~(\ref{equcon}), besides the trivial solution 
$\delta=0$, finite-$\delta$ solutions 
can be obtained from the gap equation,
\begin{equation}
1=4\lambda\int_{-\frac{\pi}{2}}^{\frac{\pi}{2}}\frac{\tanh
\left(\frac{\beta E_k}{2}\right)}{ E_k}\sin^2(k)dk,
\end{equation}
which show the typical behavior of an order parameter for a 
second-order phase transition (cf. inset of Fig. \ref{ordparplot}).

As the $Y$ and $Z$ operators depend on the phonon band index $\nu$, 
Eq.~(\ref{DMatsubara}) for the phonon Green's function must be modified. 
An additional term $C(q,i\omega_n)$ in the denominator appears. The RPA 
propagator becomes 
\begin{eqnarray}\label{2-2Matsu}
D_\nu(q,i\omega_n)&=&D^{(0)}_\nu(q,i\omega_n)\nonumber\\
&&\hspace*{-1cm} \times\frac{1-\sum\limits_{\mu\ne\nu}P_\mu(q,i\omega_n)
D^{(0)}_\mu(q,i\omega_n)}{1-\sum\limits_{\mu}P_\mu(q,i\omega_n)
D^{(0)}_\mu(q,i\omega_n)+C(q,i\omega_n)},
\end{eqnarray}
with
\begin{eqnarray}
C(q,i\omega_n)&=&D^{(0)}_0(q,i\omega_n)D^{(0)}_1(q,i\omega_n)\nonumber\\
&&\hspace*{-1.5cm}\times\left(P_1(q,i\omega_n)P_0(q,i\omega_n)-Q_1(q,i\omega_n)Q_0(q,i\omega_n)\right),
\end{eqnarray}
\begin{eqnarray}
P_\nu(q,i\omega_n)&=&g_{\nu,-q}g_{\nu,q}\bigg(\chi^\gamma_\nu(q,i\omega_n)+
\chi^\beta_\nu(q,i\omega_n)\nonumber\\
& &\hspace{1cm}+\zeta^{\beta\gamma}_\nu(q,i\omega_n)+
\zeta^{\gamma\beta}_\nu(q,i\omega_n)\bigg),\\
Q_\nu(q,i\omega_n)&=&g_{\nu,-q}g_{|1-\nu|,q}\bigg(\hat{\chi}^\gamma(q,i\omega_n)+
\hat{\chi}^\beta(q,i\omega_n)\nonumber\\
& &\hspace{1cm}+\hat{\zeta}^{\beta\gamma}(q,i\omega_n)+
\hat{\zeta}^{\gamma\beta}(q,i\omega_n)\bigg).
\end{eqnarray}
Here 
\begin{eqnarray}
\chi^\gamma_\nu(q,i\omega_n)&=&\int_{-\frac{\pi}{2}}^{\frac{\pi}{2}}
\frac{dk}{4\pi}\left(1+\alpha_{k+q}(-1)^\nu\cos(\Theta_k+\Theta_{k+q})
\right)\nonumber\\
& &\times\left(1+(-1)^\nu\cos(2k+q)\right)K^{\gamma\gamma}_{k,q}(i\omega_n),\\
\hat{\chi}^\gamma(q,i\omega_n)&=&-\int_{-\frac{\pi}{2}}^{\frac{\pi}{2}}
\frac{dk}{4\pi}\sin(\Theta_k+\Theta_{k+q})\alpha_{k+q}\nonumber\\
& &\hspace{0.5cm}\times\sin(2k+q)K^{\gamma\gamma}_{k,q}(i\omega_n)
\end{eqnarray}
denote  $Y$-$Y$ (intra-band) and
\begin{eqnarray}
\zeta^{\gamma\beta}_\nu(q,i\omega_n)&=&\int_{-\frac{\pi}{2}}^{\frac{\pi}{2}}
\frac{dk}{4\pi}\left(1-\alpha_{k+q}(-1)^\nu\cos(\Theta_k+\Theta_{k+q})\right)
\nonumber\\
& &\times\left(1+(-1)^\nu\cos(2k+q)\right)K^{\gamma\beta}_{k,q}(i\omega_n),\\
\hat{\zeta}^{\gamma\beta}(q,i\omega_n)&=&-\int_{-\frac{\pi}{2}}^{\frac{\pi}{2}}
\frac{dk}{4\pi}\sin(\Theta_k+\Theta_{k+q})\alpha_{k+q}\nonumber\\
& &\hspace{0.5cm}\times\sin(2k+q)K^{\gamma\beta}_{k,q}(i\omega_n),
\end{eqnarray}
$Z$-$Z$ (inter-band) correlation functions, 
respectively, both depending 
on a generalized Lindhard kernel
\begin{equation}
K^{\gamma\beta}_{k,q}(i\omega_n)=\frac{f^\gamma_{k+q}-f^\beta_k}{i\omega_n+ 
E^\gamma_{k+q}- E^\beta_{k}}\,.
\end{equation}
Note that Eq.~(\ref{2-2Matsu}) simplifies substantially 
for the $q=0$, $\nu=1$ mode, as for $\omega_n\ne 0$, 
$q\rightarrow 0$ all terms containing $D^{(0)}_0(q,i\omega_n)$ vanish.

\subsection{Numerical Results}

The mechanism driving the spin-Peierls phase transition is best
understood by examining the phonon dynamical structure factor.
Here we concentrate on the unstable $\pi$ mode (which is folded back
to  the $q=0$, $\nu=1$ mode in the reduced Brillouin zone). When we
calculate $S(\pi,\omega)$ for an interacting electron- (spin-) phonon 
system, we typically find a broad distribution of spectral weight. It
is clear that a phonon can only be absorbed by the fermion system, 
if its energy and momentum equal the ones of a fermionic excitation.
Therefore, in the uniform phase for $q=\pi$ we found a band 
of damped excitations for $\omega<2$. In the dimerized 
phase, there is a gap in the electronic spectrum
(cf. Fig.~\ref{ordparplot}), so damped excitations exist only in the
energy interval $2\sqrt{\lambda}\delta<\omega <2$. On the other hand,
sharp peaks in the structure factor correspond either to those (bare)
phonon modes which are outside the fermionic band and thus excluded
from scattering processes by energy conservation or to {\it new} 
quasi-particle excitations of the coupled spin-phonon system.
 
\begin{figure}[bt]
\begin{center}
\epsfxsize=0.4\textwidth
\epsffile{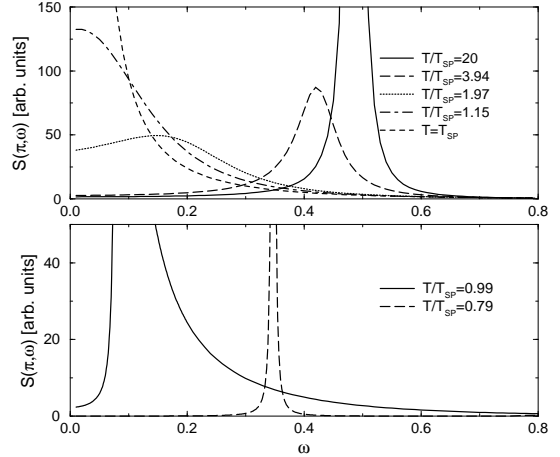}
\caption{\sl Dynamical structure factor $S(\pi,\omega)$ in the soft mode 
regime ($\Omega_\pi=0.5$, $\lambda=0.1$). In the uniform phase the high 
temperature peak softens until it reaches zero at $T=T_{\rm SP}$
(upper panel). In the dimerized phase it gets harder again (lower
panel).}\label{05Sfplot} 
\end{center}
\end{figure}

\begin{figure}[bt]
\begin{center}
\epsfxsize=0.4\textwidth
\epsffile{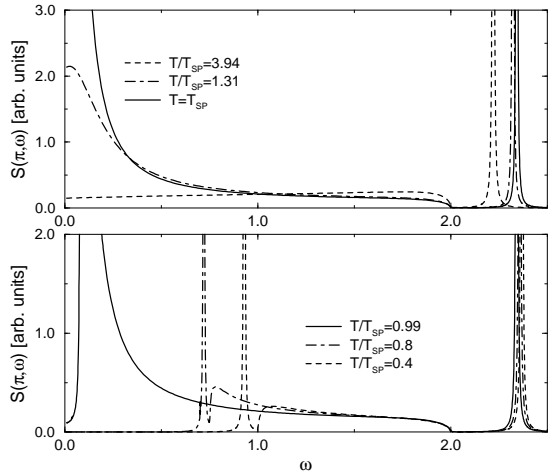}
\caption{\sl Dynamical structure factor $S(\pi,\omega)$ in the 
central peak regime 
($\Omega_\pi=2.1$, $\lambda=0.1$). The high temperature peak does not soften, 
it even gets slightly harder. A peak at $\omega=0$ appears in the uniform 
phase that becomes a singularity at $T=T_{\rm SP}$ (upper panel). In
the dimerized phase, it moves towards higher energies, corresponding
to a second excitation  (lower panel).}\label{21Sfplot}
\end{center}
\end{figure}

In the high temperature limit ($T\to \infty$), 
the correlation function $\chi(\pi,\omega)$ vanishes. 
Therefore we get a sharp peak at $\omega=\Omega_\pi$ corresponding 
to the non-interacting phonon. As the temperature is lowered 
two distinct regimes appear, depending on the frequency of the
$\pi$ phonon. For low values of $\Omega_\pi$,
i.e., in the {\it adiabatic regime}, 
the high temperature peak moves towards lower 
energies and substantially broadens, until it reaches 
zero, where it stays and gets larger in magnitude 
until a divergence appears at $T=T_{\rm SP}$ (Fig.~\ref{05Sfplot}). 
This is called a {\it soft mode scenario}. Below the 
transition, the peak moves towards higher energies. 
For large values of $\Omega_\pi$, i.e., in the {\it anti-adiabatic
regime} we found a completely different behavior, usually termed 
{\it central peak scenario}. Here the high temperature peak does not soften, 
it even gets harder. However, with lowering the temperature a maximum
in $S(\pi,\omega)$ arises at $\omega=0$, related to quasi-elastic scattering
processes. The height of this peak structure increases with decreasing
temperature until it diverges at $T=T_{\rm SP}$, where  
$S(\pi,\omega)\propto\omega^{-2}$ (Fig.~\ref{21Sfplot}). 
For $T<T_{\rm SP}$ the structure factor consists of
three parts: a delta peak slightly above $\Omega_\pi$ which can be
attributed to the original $\pi$-phonon mode, the scattering continuum 
in the range $2\sqrt{\lambda}\delta<\omega<2$ and a pronounced peak 
below the continuum, which is the central peak that moved from $\omega=0$ to 
higher energies. Let us point out that the magnetostrictive Heisenberg model
shows the same qualitative behavior. This model will be studied in
more detail in Sec.~\ref{Quasielastic}, also in relation to the
experimental findings for $\rm GuGeO_3$ at $T>T_{\rm SP}$.

Here we complete our study of the magnetostrictive XY model by examining the 
pole structure of the retarded propagator (for $q=\pi$) in 
the whole complex $\omega$-plane. The prescription (\ref{cont}) can 
easily be generalized for a complex $\omega$ in the upper 
half plane. In the lower complex plane, however, we are faced with the 
problem that $\chi(q,\omega)$ has a branch cut on the real axis at 
$\omega\in[-2;2]$ (in the uniform phase). This means that there are two 
possibilities to continue $D^{\rm ret}$ analytically to the lower half plane, 
i.e., there exist two branches. 
The first branch is analytical everywhere on the complex 
plane, except at $\omega\in [-2;2]$, the second everywhere except 
$\omega\in]-\infty;-2]\cup[2;\infty[$. They are both of course identical in 
the upper half plane. The first branch is directly obtained by evaluating the 
integral in Eq.~(\ref{dimdimcor}) for an $\omega$ 
with $\textrm{Im}\omega<0$. We get the 
second by extrapolation from the upper half plane for 
$\textrm{Re}\omega\in[-2;2]$ (this is done by a fourth order power series). 
It turns out that the first branch will yield purely real poles with 
$\textrm{Re}\omega>2$, the second branch corresponds to poles 
with negative imaginary part and $\textrm{Re}\omega<2$.

\begin{figure}[bt]
\begin{center}
\epsfxsize=0.4\textwidth
\epsffile{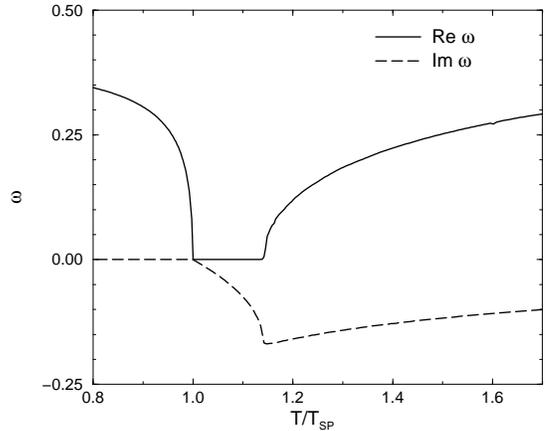}
\caption{\sl Real and imaginary part of the pole of the retarded Green's 
function in the soft mode regime ($\Omega_\pi=0.5$, $\lambda=0.1$).}
\label{05poleplot}
\end{center}
\end{figure}

\begin{figure}[bt]
\begin{center}
\epsfxsize=0.4\textwidth
\epsffile{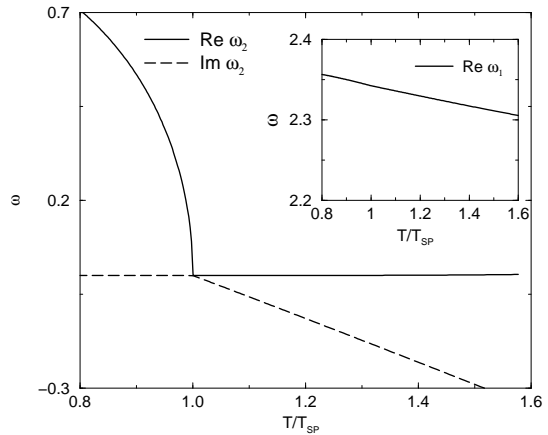}
\caption{\sl Poles of the retarded Green's function in the central peak 
regime ($\Omega_\pi=2.1$, $\lambda=0.1$). The (real) high temperature pole 
$\omega_1$ does no soften, as shown in the inset. Instead a second pole 
$\omega_2$ appears, which is purely imaginary in the uniform phase and gets 
real in the dimerized phase.}\label{21poleplot}
\end{center}
\end{figure}

When lowering the temperature in the {\it soft mode regime} (see
Fig.~\ref{05poleplot}) the real part of the high temperature pole
decreases from $\Omega_\pi$ to $0$ at a temperature larger than
$T_{\rm SP}$. The modulus of the imaginary part grows with decreasing
temperatures and reaches a maximum at the temperature where the real
part gets to zero for the first time. Then it decreases again. At
$T=T_{\rm SP}$ we have $\omega=0$ as expected. For lower temperatures the
pole is real and its value increases until it saturates at $T=0$. 
In the {\it central peak regime} (see Fig.~\ref{21poleplot}) 
the high temperature pole $\omega_1$ gets harder. 
A second pole $\omega_2$, which is purely imaginary 
for $T>T_{\rm SP}$ appears and causes the instability at $T=T_{\rm SP}$. For 
$T<T_{\rm SP}$ this pole is real and increasing as $T\to 0$. 

Although the interpretation of real and imaginary part of a complex 
singularity as energy and damping of a quasiparticle is doubtful in the 
presence of a branch cut,\cite{EngelsbergSchrieffer63} examining the pole 
structure still gives a qualitative understanding of the mechanism driving 
the phase transition. Most notably, the purely imaginary structure factor 
just above $T_{\rm SP}$ signals quasi-elastic scattering, i.e. the 
existence of diffusive modes, in both the adiabatic and anti-adiabatic 
regimes. Of course, to get a complete picture, 
it is also necessary to take into account the spectral weight 
of the continuum seen in the structure factor.

\subsection{Application to CuGeO$_3$}\label{Application}

Even though the model we discussed is far too simple to expect 
a good quantitative description of CuGeO$_3$ 
it should at least produce effects in the right 
order of magnitude. To make contact with the experimentally observed
magneto-elastic excitation spectrum of CuGeO$_3$,
in the numerical calculations we fix the energy scale
by  $J=150$~K and use $T_{\rm SP}=14$ K together with a phonon frequency 
of $\Omega_{{\bf q}_0}/2\pi=6.53$ THz which corresponds to the
dominant Peierls-active $T_2^+$-phonon-mode.\cite{Werner99CGO} 
This gives the control parameters $\Omega_\pi=2.09$ and 
$\lambda=0.057$. For the spin gap at $T=0$ we get 
$2\Delta=2\sqrt{\lambda}\delta J=4.422$ meV compared to an experimental value
of $2\Delta=4.2$ meV.\cite{Els97} The ground state exchange alternation is 
$\delta_J=\sqrt{\lambda}\delta=0.1645$.  Other methods give values from $0.01$
 to $0.2$.\cite{Werner99CGO} 

\begin{figure}[bt]
\begin{center}
\epsfxsize=0.4\textwidth
\epsffile{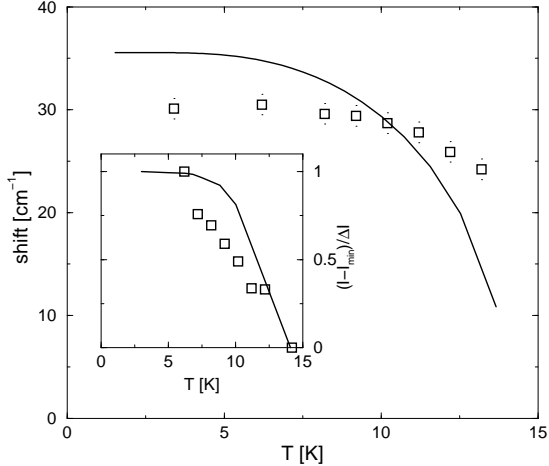}
\caption{\sl Temperature dependence of the energy of the second 
magneto-elastic excitation obtained from the magnetostrictive XY model 
(solid line) compared to experimental data for the $30$ cm$^{-1}$ mode 
of CuGeO$_3$ (symbols)~\protect\cite{Els97}. 
The inset shows the rescaled spectral weight 
of this excitation vs. temperature (solid line) 
compared to experiment (symbols).~\protect\cite{Lemmens}}
\label{30cmplot}
\end{center}
\end{figure}

As expected for CuGeO$_3$, we are in the central peak regime.  
Thus we have a second excitation for $T\le T_{\rm SP}$. In recent
inelastic light scattering (ILS) experiments a peak in the spectrum at 
$30$ cm$^{-1}$ was observed and interpreted as a singlet 
bound state of two antiparallel magnons.\cite{Els97} 
One could now speculate that the excitation below 
scattering continuum we found in the structure factor 
is the phonon contribution of this new magneto-elastic excitation. 
A comparison of the theoretical and experimental\cite{Lemmens} 
data for the position and intensity of this peak is given in 
Fig.~\ref{30cmplot}. Theory and experiment show the same overall behavior, 
although the decrease of the peak position is much more pronounced in the 
theory.

Here the spin-phonon coupling gives rise to an effective spin-spin 
interaction\cite{WWF99} which, in the dimerized phase, 
leads to a {\it phonon-induced} bound state in the magnetic 
excitation spectrum just below the fermionic scattering continuum.  
A signature of this bound state appears in the phonon structure factor 
as shown in Fig.~\ref{21Sfplot} proving its magneto-elastic character. 
In the case of dimerized Heisenberg chains the Jordan-Wigner transformation 
of the spin part gives directly a four-fermion interaction 
term\cite{Fradkin} leading to a peak in the dimer-dimer 
correlation function at $\sqrt{3}/{2}$ times the band gap.\cite{US96,ETD97} 
Due to the phonon-induced effective spin-spin (fermion-fermion) interaction,
the energy of this bound state will be shifted in the order of 
$\sum_\nu|g_{\nu,{\bf q}_0}|^2/(\hbar\Omega_{\nu,{\bf q}_0}) \approx
0.1J$,\cite{WernerThesis} which is a 10 \% effect, 
and the bound state will appear in the phononic structure factor. 
This supports the interpretation of the peak at $30$ cm$^{-1}$ 
observed with inelastic light scattering 
as the singlet bound state as shown in Fig.~\ref{30cmplot}. 
The quasi-elastic scattering is the precursor of this excitation. 
Moreover, the hardening of in the Peierls-active phonon modes  
observed in CuGeO$_3$\cite{Braden98CGO} is qualitatively well 
described within the RPA scheme,\cite{Gros98CGO,Werner00ED}
i.e., those initial elastic excitations also have a magnetic character.

To summarize, in this section we have performed a comprehensive study of
the magnetostrictive XY model which is the minimal model 
capable of  describing the spin-Peierls scenario in the whole 
phonon frequency range. The focus was on the 
anti-adiabatic central-peak regime being relevant for CuGeO$_3$. 



\section{Quasi-elastic scattering in
C\lowercase{u}G\lowercase{e}O$_3$}\label{Quasielastic}

In order to describe experimental results on CuGeO$_3$ accurately,
the spin system to include are frustrated Heisenberg
chains\cite{Hase93CGO,Castilla95CGO,Riera95CGO,FKL+98} which is
coupled to the four Peierls-active phonon modes. Frequencies and
coupling constants of the Peierls-active T$_2^+$ phonons are discussed
in detail in Refs.\ \onlinecite{Braden98CGO} and
\onlinecite{Werner99CGO}. The CuGeO$_3$ samples used for comparison
with experiment in this section undergo the spin-Peierls transition at
$T=14.3$ K, the theoretical calculations are adapted to match via the 
coupling constants. The wave vector of the modulation in the ordered
phase is ${\bf q}_0=(\pi/a,0,\pi/c$), where $a$ and $c$ are the
lattice constant along the crystallographic $x$ and $z$ direction, 
respectively. We set $J/k_{\rm B}=150$ K, which is together with a
value of $J_2/J=0.24$ among those discussed as valid for
CuGeO$_3$.\cite{Werner99CGO,Brenig97CGO} Quasi-elastic scattering has
been observed in neutron scattering
experiments\cite{Hirota95CGO,BradenHabil} up to 16 K and in X-ray
scattering\cite{Pouget96CGO} up to 40 K or $k_{\rm B}T\approx 0.3J$.
The constants $k_{\rm B}$ and $\hbar$ are explicitly given in this
section for a more transparent unit conversion.

The dimer-dimer correlation function as given in Eq.\ (\ref{chi0})
has been calculated for Heisenberg chains in the uniform phase by
Cross and Fisher\cite{CF79} with bosonization techniques. In the
analytically continued form one has
\begin{eqnarray}\label{chiCF}
\chi_{\rbx{\rm CF}}(q_z,\omega)&=& 
\nonumber\\&&\hspace{-9ex}
       \frac{-\chi_{\rbx{0}}(\frac{k_{\rm B}T}{J})}{0.35\,k_{\rm B}T}\ 
	I_1\left[\frac{\omega - v_s |q_z - \frac{\pi}{c}|}
{2\pi (k_{\rm B}/\hbar) T}\right] \,
	I_1\left[\frac{\omega + v_s |q_z - \frac{\pi}{c}|}
{2\pi (k_{\rm B}/\hbar) T}\right]
\nonumber\\
\end{eqnarray}
with the spin-wave velocity\cite{Fledder97CGO}
$v_s=c(J-1.12J_2)/(\pi\hbar)$ and the functions  
$
I_1(k)=(8\pi)^{-1/2}\, \Gamma(\frac{1}{4}-\frac{1}{2}ik)\, 
	\Gamma^{-1}(\frac{3}{4}-\frac{1}{2}ik)$.
The result has the general form of spin correlation functions obtained
from conformal field theory.\cite{Tsvelik,Schulz86LL,BK81} The choice
of the value of $J_2=0.24J \le J_c$ allows for the application of the
field-theoretical results. For $J_2>J_c$ the spectrum of the spin
system is gaped.\cite{Chitra95DMRG}

The prefactor $\chi_{\rbx{0}}(k_{\rm B}T/J)$ is assumed constant in
field theory but has been shown by Raupach {\it et al.}\ using density
matrix renormalization group (DMRG) studies to be temperature
dependent in the static case and for $q_z=\pi/c$.\cite{Raupach99CGO}
Recent numerical studies suggest that the approximate result Eq.\
(\ref{chiCF}) describes the exact dimer-dimer correlation function
better when rescaling the energy as
$
\chi (q_z,\omega)=
\chi_{\rbx{\rm CF}}(q_z,g_T\,\omega)$,
where the scaling function $g_T$ depends on the NNN coupling $J_2$. We
use in this work the $g_T$ as given in Ref.\ \onlinecite{Werner00ED}
for $J_2/J=0.24$. Please note that none of the results presented
herein depend qualitatively on the inclusion of $g_T$.

\subsection{Frequency dependence}

\begin{figure}[bt]
   \epsfxsize=0.50\textwidth
   \centerline{\epsffile{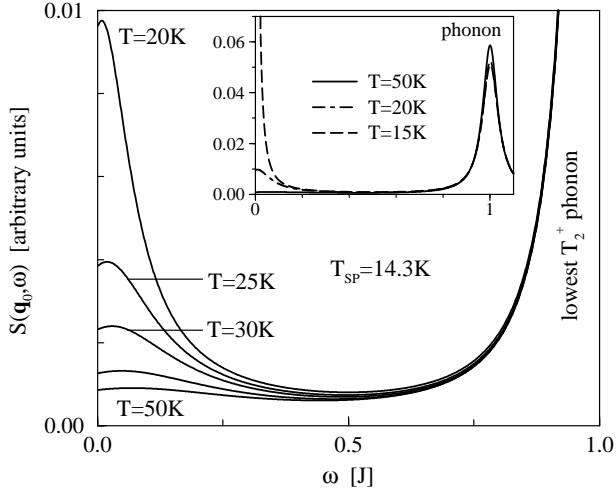}}
   \centerline{\parbox{\textwidth}{\caption{\label{Svonomega}
   \sl Frequency dependence of the structure factor from Eq.\
   (\protect\ref{dynamical}) appropriate for CuGeO$_3$. As the
   temperature approaches the spin-Peierls transition quasi-elastic
   scattering appears. Inset: larger scale representation. The phonon
   is the lowest of the four Peierls-active modes, see Refs.\
   \protect\onlinecite{Braden98CGO,Werner99CGO}.  
   }}}
   \end{figure}

The frequency dependence of the dynamical structure factor Eq.\
(\ref{dynamical}) for the parameters relevant for CuGeO$_3$ is shown
in Fig.\ \ref{Svonomega}. Below $T\sim 3T_{\rm SP}$ spectral weight
appears in the center of the spectrum. Expanding the complex
$\chi_{\rbx{\rm CF}}(\pi/c,\omega)$ to second order\cite{WernerThesis}
in $\omega$ in Eq.\ (\ref{DMatsubara}), the dynamical structure factor
can be determined for $\hbar\omega\ll k_{\rm B}T_{\rm SP}$ to diverge
as $S({\bf q}_0,\omega)|_{T=T_{\rm SP}}\sim\omega^{-2}$ at the phase
transition.

By analogy to the XY model discussed in the previous section (compare
Figs.\ \ref{21Sfplot}, \ref{21poleplot}, and Sec.\ \ref{Application})
this is the precursor of the magneto-elastic mode appearing at the
phase transition. The Peierls-active phonons (only the lowest
is shown in Fig.\ \ref{Svonomega}) harden as the temperature is
lowered and the temperature dependence of the intensity of the
quasi-elastic scattering is consistent with neutron- and
X-ray-scattering experiments.\cite{Werner00ED}

\subsection{Momentum dependence}\label{momentumsection}

The momentum dependent scattering rate of inelastic neutrons or X-rays
can be obtained by convoluting the dynamical structure factor Eq.\
(\ref{dynamical}) with a Gaussian of the width of the experimental
energy resolution $\sigma_\omega$ and a Gaussian of the width of the
experimental momentum resolution $\sigma_{q_z}$. The limitation to the
chain direction is imposed since the dimer-dimer correlation function
Eq.\ (\ref{chiCF}) only introduces a $z$ axis dispersion:
\begin{eqnarray}\label{Intensity}
I(q_z,\omega)&=&\frac{1}{2\pi\sigma_\omega\sigma_{q_z}}
      \int\limits_{-\infty}^\infty d\omega'\, 
	{\rm e}^{-\frac{(\omega'-\omega)^2}{2\sigma_\omega^2}}\ 
      \int\limits_{\rm 1.BZ} d^3 q \,
	{\rm e}^{-\frac{(q_z'-q_z)^2}{2\sigma_{q_z}^2}}
\nonumber\\ && \times\
        \delta(q_x-\pi/a)\ \delta(q_y)\
      S({\bf q}',\omega')\,.
\end{eqnarray}
The first Brillouin zone (1.BZ) is that in the disordered high
temperature phase. 

\begin{figure}[bt]
   \epsfxsize=0.50\textwidth
   \centerline{\epsffile{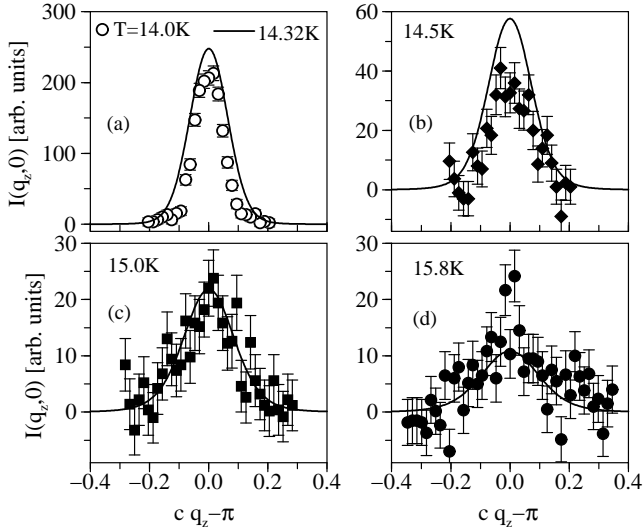}}
   \centerline{\parbox{\textwidth}{\caption{\label{Iofq}
   \sl Quasi-elastic scattering from neutrons (symbols, background
   subtracted, scaled with a unique factor for all temperatures)
   compared with theory from Eq.\ (\protect\ref{Intensity}) (full
   lines) for different temperatures. (a) Due to critical fluctuations
   the experimental intensity does not diverge at the transition and 
   reaches the value of the theoretical results only somewhat below
   the transition. In (b) temperature is still in the critical region,
   (c) and (d) show excellent agreement. The experimental scans run
   along $[\pi/a(1+2p), \pi/b(8-2p), \pi/c(1+2p)]$, see Sec.\
   \protect\ref{momentumsection}.	
   }}}
   \end{figure}

Fig.\ \ref{Iofq} shows plots of $I(q_z,0)$ for different temperatures
in comparison with neutron scattering data from Ref.\
\onlinecite{BradenHabil}. Parameters are chosen appropriate for CuGeO$_3$
as discussed above, the resolutions are
$\hbar\sigma_\omega\approx0.05J$ and $\sigma_{q_z}\approx
0.06/c$. The value of $\sigma_\omega$ is given by the experimental
setup,\cite{Gros98CGO} $\sigma_{q_z}$ is obtained from the resolution
limited Bragg peak at 4 K.\cite{BradenHabil} The agreement with
experiment is satisfactory. Note that the critical region has been
estimated via the Ginzburg criterion\cite{Werner99CGO} to be $T_{\rm
SP}\pm 0.4$ K. Within this region the theoretical divergence of the
intensity\cite{Werner00ED} is suppressed by critical fluctuations.

The experimental scans run along $[\pi/a(1+2p), \pi/b(8-2p),
\pi/c(1+2p)]$. $p$ is the running parameter, $q_y=8\pi/b$ assures that
there is no significant magnetic contribution to the
signal.\cite{BradenHabil} Since $c<a<b$ and since the correlations
along $q_z$ are clearly dominant,\cite{Hirota95CGO,Pouget96CGO} the
data are still eligible for comparison with the theoretical data along
$q_z$.

\subsection{Correlation length}

When correcting the momentum dependent quasi-elastic scattering
shown in Fig.\ \ref{Iofq} for experimental resolution the data can be
fitted nicely by Lorentzians.\cite{Hirota95CGO,WKC+00} For $T-T_{\rm
SP}\le0.1$ K a second length scale appears which can be fitted by an
additional Lorentzian squared contribution. It is attributed to
surface strain effects\cite{WKC+00,Harris95CGO} not included in the
RPA treatment. 

\begin{figure}[bt]
   \epsfxsize=0.50\textwidth
   \centerline{\epsffile{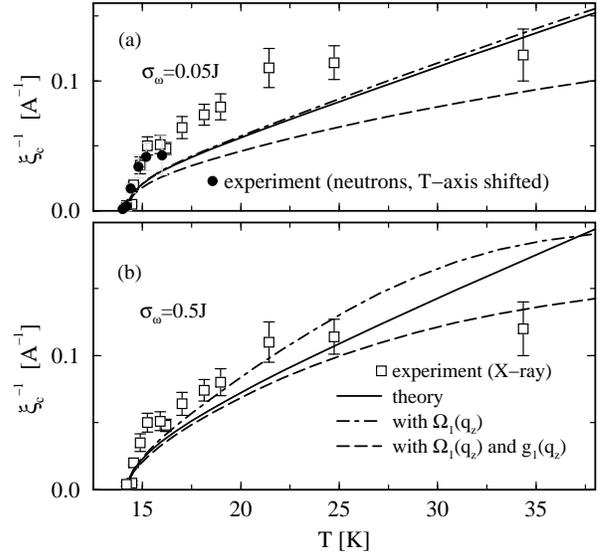}}
   \centerline{\parbox{\textwidth}{\caption{\label{Korrlen}
   \sl Inverse correlation length along the magnetic chains from RPA
   compared with experimental data from Refs.\
   \protect\onlinecite{Pouget96CGO} (squares) and
   \protect\onlinecite{Hirota95CGO} (circles).  The lower resolution
   curves (a) are lower than those with larger $\sigma$ modeling the
   X-ray experiment (b). The larger $\sigma$ values are sensitive to
   the inclusion of the phonon dispersion into theory (dash-dotted
   lines). Including a Lorentzian shape of the spin-phonon coupling
   constants as a function of (q) leads to a saturation of $1/\xi$ at
   lower values (dashed lines).
   }}}
   \end{figure}

From the width of the fits the correlation length can be
extracted. The corresponding theoretical correlation length is
obtained accordingly from Eq.\ (\ref{Intensity}) setting
$\sigma_{q_z}=0$. Figure \ref{Korrlen} shows the importance of the
energy resolution when discussing the temperature dependence of the
extracted correlation length. The symbols mark the experimental data,
open squares are from X-ray data in Ref.\ \onlinecite{Pouget96CGO},
full circles are neutron data in Ref.\ \onlinecite{Hirota95CGO}. The
neutron data are shifted by $\Delta T=1$ K in order to match the
different critical temperatures of the samples. 

The energy resolution in neutron scattering is of the order
of a few meV while X-rays integrate over a much larger energy
interval. $\hbar\sigma_\omega\approx0.05J$ simulates the resolution of
diffracted neutrons\cite{Gros98CGO} and $\hbar\sigma_\omega\approx 0.5J$ is
relevant for X-ray scattering. The X-ray resolution is probably even
larger, but the interval of $-0.5J<\hbar\omega<0.5J$ covers the full width
of the relevant magnetic spectrum.\cite{Werner00ED} 

The following conclusions can be drawn from the results presented:

(i) The energy integration in X-ray scattering allows for the
determination of the correlation to much higher temperatures than
neutron scattering.\cite{Werner00ED} 

(ii) The momentum and frequency dependence of $S({\bf q},\omega)$
cannot be factorized yielding the different magnitude of the
theoretical results for $\hbar\sigma_\omega = 0.05J$ and
$\hbar\sigma_\omega = 0.5J$, shown by the full lines in Fig.\
\ref{Korrlen} (a) and (b), respectively. 

(iii) The large energy integration makes the X-ray results sensitive to
phonon dispersion effects. The results in Ref.\
\onlinecite{Braden98CGO} suggest a small dispersion for the lowest
Peierls active phonon which we model by $\Omega_{1}(q_z) \approx
\Omega_{1,{\bf q}_0} (1 + 0.7\,|c\,q_z-\pi|^2)$ for $|c\,q_z-\pi|\ll
1$ with $\hbar\Omega_{1,{\bf q}_0}\approx J$. While for a resolution
of $\hbar\sigma_\omega = 0.05J$ the correlation length is basically
independent of the dispersion (dash-dotted line in Fig.\ \ref{Korrlen}
(a)), the $\hbar\sigma_\omega = 0.5J$ data clearly are altered
(dash-dotted line in Fig.\ \ref{Korrlen} (b)).

(iv) The coupling constants $g_{\nu,{\bf q}}$ in Eq.\ (\ref{Hspbose})
depend on the polarization vectors of the phonon
modes.\cite{Werner99CGO} Dephazation effects suggest a suppression of
$g_{\nu,{\bf q}}$ away from ${\bf q}_0$. We model this suppression in
the normal coordinate propagator Eq.\ (\ref{DMatsubara}) by a
Lorentzian along $q_z$
\begin{equation}\label{gvonq}
g_{\nu,{\bf q}}g_{\nu,-{\bf q}}\approx \frac{1 - \cos(q_z c)}{2}\
   \frac{(c/\kappa)^{2}}{(c/\kappa)^{2} + 
         (q_z c - \pi)^2}\ |g_{\nu,{\bf q}_0}|^2\,.
\end{equation}
The full and dash-dotted curves in Fig.\ \ref{Korrlen} are obtained in
the limit $c/\kappa\to \infty$. Setting $c/\kappa=0.5$ yields
the dashed curves in Fig.\ \ref{Korrlen} (a) and (b). The saturation
of the correlation length at higher temperatures suggests a value of 
$c/\kappa < 0.5$ for CuGeO$_3$.

(v) Close to the phase transition neither the phonon dispersion nor
the coupling constant's $q_z$ dependence are important. The RPA
results yield $\xi_c \sim (T-T_{\rm SP})^{0.5}$ in agreement with
detailed X-ray investigations by Harris {\it et al.}\cite{Harris95CGO}

(vi) The curves obtained for $J_2=0$ (not shown) lie about 30\% below
those for $J_2/J=0.24$ (Fig.\ \ref{Korrlen}) suggesting that
$\alpha=J_2/J\ge 0.24$ for CuGeO$_3$.\cite{Werner99CGO,Brenig97CGO}

The overall agreement of the description of the experimental data on
the quasi-elastic scattering in CuGeO$_3$ is quite satisfactory. This
suggests that indeed it is the precursor of the magneto-elastic mode
discussed in the previous section.

\section{Real space interpretation}

An open question is the appropriate real-space interpretation of the
quasi-elastic scattering, especially in the central peak regime. To
obtain a qualitative picture we consider the effective action in RPA
as is was derived for example in Ref.\ \onlinecite{WernerThesis}. We
limit ourselves here to a single phonon mode and only consider the
physics within a chain of magnetic Cu ions. The action then reads in
the static limit
\begin{eqnarray}\label{ActionRPA}
{\cal S}_{\rm RPA} &=&
\beta \sum_{q_z}
              \hbar\Omega_{q_z} |\phi^{}_{q_z}|^2       
\nonumber\\&+&
 \beta \sum_{q_z} \frac{g_{q_z}g_{-q_z}}{2}\ \chi_{\rbx{\rm CF}}(-q_z,0)\
         \left|\phi^*_{-q_z} + \phi^{}_{q_z}\right|^2 .
\end{eqnarray}
The phonon fields $\phi^{}_{q_z}$ and $\phi^{*}_{q_z}$ are directly
related to the Bose operators $b^{}_{q}$ and $b^{\dagger}_{q}$ in the
minimal model discussed in Sec.\ \ref{minimal}. 

The first term of Eq.\ (\ref{ActionRPA}) simply corresponds to the
phonon Hamiltonian $H_p$ in Eq.\ (\ref{Hminimal}) of the minimal
model. The second term of Eq.\ (\ref{ActionRPA}) is the relevant
correction term. Within the approximations made and neglecting the
momentum dependence of the polarization vector the coupling constants
are given by 
\begin{equation}\label{gofq}
\frac{g_{q_z}g_{-q_z}}{2}=[1-\cos(q_zc)]\ \left(g^z_{\rm\tiny Cu}\right)^2
         \ \frac{\hbar}{2\Omega_{q_z}m_{\rm\tiny Cu}},
\end{equation}
$g^z_{\rm\tiny Cu}$ is the change of $J$ with the Cu elongation, and
$m_{\rm\tiny Cu}$ is the Cu mass.\cite{Werner99CGO}

The appropriate transformation of the reciprocal-space fields
$\phi^{*}_{q_z}$ to real-space elongation fields $u_{l_z}$ is
\begin{equation}\label{phitou}
\phi^*_{-q_z} + \phi^{}_{q_z} = 
     \sqrt{\frac{2\Omega_{q_z}m_{\rm\tiny Cu}}{\hbar L}}
     \sum_{l_z} {\rm e}^{-iq_zcl_z}\ u_{l_z},
\end{equation}
the conjugated momenta are given by
\begin{equation}\label{phitop}
\phi^*_{-q_z} - \phi^{}_{q_z} = \frac{1}{i}\
     \sqrt{\frac{1}{2\hbar\Omega_{q_z}m_{\rm\tiny Cu} L}}
     \sum_{l_z} {\rm e}^{-iq_zcl_z}\ p_{l_z}.
\end{equation}
The $q_z$ dependence of the dimer-dimer correlation function
$\chi_{\rbx{\rm CF}}(q_z,0)$ is satisfactorily approximated by a
Lorentzian
\begin{equation}\label{Chifit}
\chi_{\rbx{\rm CF}}(q_z,0)\approx
      \frac{-\chi_{\rbx{0}}(\frac{k_{\rm B}T}{J})}{k_{\rm B}T}\
      \frac{(c/\xi_D)^2}{(c/\xi_D)^2 + (q_zc - \pi)^2}\,. 
\end{equation}
The dimer length scale was defined as
\begin{equation}\label{dimerlength}
\xi_D^{-1} = 0.53\ \frac{2\pi}{\hbar v_s}\ k_{\rm B}T\,.
\end{equation}

\begin{figure}[bt]
   \epsfxsize=0.50\textwidth
   \centerline{\epsffile{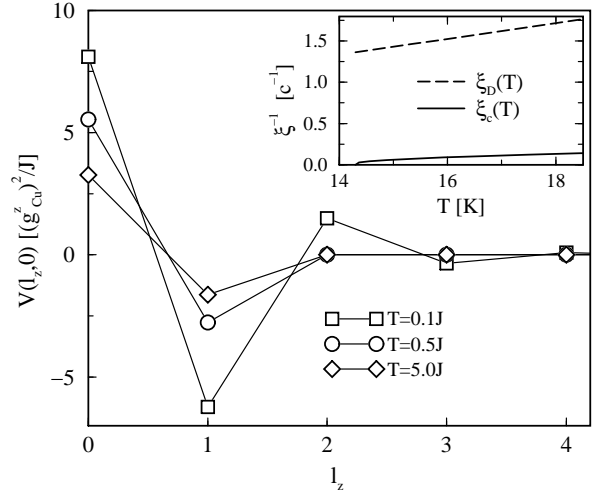}}
   \centerline{\parbox{\textwidth}{\caption{\label{VvonT}
   \sl Spin-phonon coupling induced dimerization potential as given in
   Eqs.\ (\protect\ref{Vdimer}) and (\protect\ref{Vnull}) for
   different temperatures and $J_2/J=0.24$. The inset
   shows the inverse dimer correlation length as in Eq.\
   (\protect\ref{dimerlength}), dashed line, in comparison with the
   spin-Peierls correlation length, full line identical to the full
   line in Fig.\ \protect\ref{Korrlen} (a).
   }}}
   \end{figure}

Introducing the dispersion as given in Eq.\ (\ref{1Ddisp}) and
applying Eqs.\ (\ref{gofq}) through (\ref{dimerlength}) to Eq.\ 
(\ref{ActionRPA}) the action becomes  
\begin{eqnarray}\label{ActionReal}
{\cal S}_{\rm RPA} &\approx&
\beta \sum_{l_z} 
     \frac{p_l^2}{2 m_{\rm\tiny Cu}} +
     \frac{\Omega_\pi^2 m_{\rm\tiny Cu}}{8}(u_l - u_{l+1})^2      
\nonumber\\&-&
 \beta \sum_{l_z,l'_z}\ 
         V(l_z,l'_z)\ u_{l_z}u_{l'_z}\,.
\end{eqnarray}
The dimer-dimer correlation induced potential 
\begin{eqnarray}\label{Vdimer}
V(l_z,l'_z) &=& V_0\bigg[\delta_{l_z=l'_z} +
\frac{\delta_{l_z\neq l'_z}}{2}\ (-1)^{|l_z-l'_z|}
\nonumber\\&&\hspace{12ex}\times      
      \ {\rm e}^{-\frac{c|l_z-l'_z|}{\xi_D}}\ 
      (1+\cosh \frac{c}{\xi_D})\bigg]
\end{eqnarray}
with amplitude
\begin{equation}\label{Vnull}
V_0=\left(g^z_{\rm\tiny Cu}\right)^2\ \frac{1.06\pi c}{\hbar v_s}\
         \chi_{\rbx{0}}(\frac{k_{\rm B}T}{J})
\end{equation}
is alternating in space and decaying on the length scale of
$\xi_D\sim T^{-1}$. The amplitude of the potential is determined by
$\chi_{\rbx{0}}(k_{\rm B}T/J)$ of which the temperature dependence is
shown in Ref.\ \onlinecite{Raupach99CGO}. It is enhanced for
$T<J/k_{\rm B}$ and appears to vanish for $T\to 0$ for $J_2=0$ while
it might even diverge for $J_2/J\ge 0.241$. The potential $V(l_z,0)$
is plotted for different temperature for $J_2/J = 0.24$ in Fig.\
\ref{VvonT}. Note that $V(l_z,l'_z)$ is translational invariant.

This potential enhances local dimerization on the length scale
$\xi_D$. It is crucial to distinguish $\xi_c\neq\xi_D$ (see inset of Fig.\
\ref{VvonT}). The correlation length of the spin-Peierls transition
$\xi_c$ describes fluctuations to be associated with the coherent
three-dimensional ordering of the local dimerized areas of scale
$\xi_D$. This coherent dimer-ordering in CuGeO$_3$ has been 
described very successfully via effective Ising based mean-field
models.\cite{WernerThesis,BKW99,BNR98,Werner98CGO} In this sense the
spin-Peierls transition in CuGeO$_3$ can be considered as a
order-disorder transition where the objects that order are only
induced by the spin system as the temperature is lowered substantially
below $J/k_{\rm B}$. The coherent ordering leads to tricritical
behavior\cite{BKW99} with a tricritical to mean-field crossover
temperature of $T_{\rm CR} - T_{\rm SP} \approx 0.1$ K coinciding with
the appearance of large length scale fluctuations.\cite{WKC+00} 

A straightforward determination of the magnetic dimer correlation
length in Eq.\ (\ref{dimerlength}) with the parameters for CuGeO$_3$
as discussed above yields $\xi_D(T_{\rm SP})\approx 0.7c$. Considering
the momentum dependence of the coupling constants as discussed in Eq.\
(\ref{gvonq}) rescales the magnetic dimer correlation length roughly
as $\overline{\xi}_D\approx\sqrt{\xi_D^2 + \kappa^2}$. For
$c/\kappa=0.5$ one has $\overline{\xi}_D(T_{\rm SP})\approx
2c$ which then is basically temperature independent. The order of
magnitude is reasonable.

Note that this real-space interpretation is consistent with the
pre-transitional pseudo-gap behaviour discussed in the context of
Peierls transitions.\cite{Lee73Peierls,McKenzie95I,McKenzie95II}

\section{Conclusions}

We discussed in detail magneto-elastic excitations in systems of
phonons coupled to spin chains within the random phase
approximation. The XY model allowed for an exact determination of the
temperature dependence of the poles of the dynamical structure factor
in the disordered as well as in the dimerized phase in both the soft
phonon and the central peak regime. The model of frustrated
Heisenberg chains coupled to phonons applied to the spin-Peierls
system CuGeO$_3$ correctly describes the details of the quasi-elastic
scattering such as its frequency dependence, momentum space
dependence, and the extracted correlation lengths. The importance of
the experimental energy resolution is emphasized.

The quasi-elastic scattering can be interpreted as the precursor of a
new magneto-elastic excitation in the dynamical phonon structure
factor for $T<T_{\rm SP}$ which increasingly splits off from the
scattering continuum as the temperature is lowered. In alternating
Heisenberg chains relevant for CuGeO$_3$ this leads to a
renomalization of the singlet bound state by about 10 \%. In
alternating XY chains the position and temperature dependence of this
excitation can be calculated explicitly and compares favorably with
the 30 cm$^{-1}$ mode found in inelastic light scattering experiments,
thus yielding a qualitatively modified explanation of this signal.   

The real-space interpretation of a spin-phonon induced alternating
elastic potential supports the applicability of Ising-like approaches
to the spin Peierls transition. The spin-phonon induced alternating
elastic potential driving the transition underlines the mixed
magneto-elastic character of quasi-elastic scattering.

\section{Acknowledgments}

We are indebted to M.\ Braden for furnishing the neutron scattering  
data files and instructive discussions. We thank P. Lemmens for
providing ILS data for the $30$ cm$^{-1}$ mode and discussions. We
thank A.\ P.\ Kampf, A.\ Wei{\ss}e, and A.\ Zheludev for stimulating 
discussions. The work performed in Bayreuth and in Wuppertal was
supported by DFG program ``Schwerpunkt 1073'', the work at BNL was
supported by DOE contract number DE-AC02-98CH10886.

\end{document}